\documentclass[11pt,twoside]{article}
\usepackage{asp2010}
\usepackage{chapterbib}

\resetcounters

\begin{document}

\title{Simulated ALMA observations of collapsing low-mass dense cores}
\author{F. Levrier,$^1$ B. Commer\c{c}on,$^{1,3}$ A. J. Maury,$^2$ Th. Henning,$^3$ R. Launhardt,$^3$ and C. Dullemond$^4$
\affil{$^1$LERMA/LRA - UMR 8112 - Ecole Normale Sup\'erieure, 24 rue Lhomond, 75231 Paris CEDEX 05, France}
\affil{$^2$ESO, Karl Schwarzschild Stra{\ss}e 2, 85748, Garching bei M\"unchen, Germany}
\affil{$^3$Max Planck Institut f\"ur Astronomie, K\"onigsthul 17, 69117 Heidelberg, Germany}
\affil{$^4$Zentrum f\"ur Astronomie der Universit\"at Heidelberg, Institut f\"ur Theoretische Astrophysik, Albert-Ueberle-Stra{\ss}e 2, 69120 Heidelberg, Germany}}

\begin{abstract}
We present a possible identification strategy for first hydrostatic core (FHSC) candidates and make predictions of ALMA dust continuum emission maps from these objects. We analyze the results given by the different bands and array configurations and identify which combinations of the two represent our best chance of solving the fragmentation issue in these objects. If the magnetic field is playing a role, the emission pattern will show evidence of a pseudo-disk and even of a magnetically driven outflow, which pure hydrodynamical calculations cannot reproduce.
%
\end{abstract}

\section{Radiation-magneto-hydrodynamic (RMHD) models of core collapse}
Models of protostellar core collapse are built using the adaptive-mesh refinement (AMR) code RAMSES \citep{2002A&A...385..337T}, which integrates the equations of ideal MHD \citep{2006A&A...457..371F} and the equations of radiation hydrodynamics under the gray flux-limited diffusion approximation \citep{2011A&A...529A..35C}. A $1~\mathrm{M}_\odot$ sphere of gas with initial radius 3300 AU and uniform density and temperature is put into solid body rotation \citep{2010A&A...510L...3C} around the same axis as the initial magnetic field. Three models with effective resolution of 0.2 AU are presented in \cite{2012A&A...545A..98C}, corresponding to three levels of magnetization expressed in terms of the mass-to-flux to critical mass-to-flux ratio $\mu=2$ (strong magnetization), $\mu=10$ (intermediate magnetization), and $\mu=200$ (quasi-hydrodynamical case). 

\section{Identification of FHSC candidates from spectral energy distributions (SED)}
Spectral energy distributions (SED) of dust thermal emission from the RMHD models are computed \citep{2012A&A...545A..98C} using the RADMC-3D code by C. Dullemond\footnote{\tt http://www.ita.uni-heidelberg.de/\textasciitilde{}dullemond/software/radmc-3d/}, assuming a distance of 150 pc, integrating over a 3000 AU by 3000 AU area, and at three different inclination angles. The strongly magnetized and quasi-hydro pole-on models show a clear and similar SED evolution as collapse proceeds, with a significant amount of flux emitted in the later stages between $20~\mu\mathrm{m}$ and $100~\mu\mathrm{m}$, because envelopes are not thick enough to fully reprocess the central objectÕs radiation. This is in contradiction to previous studies assuming spherical symmetry \citep[e.g.][]{2007PASJ...59..589O}. 

\section{Simulated ALMA observations of the dust thermal continuum}
Point-source detections below $100~\mu\mathrm{m}$ combined with non-detections below $10~\mu\mathrm{m}$ may thus help identify FHSC candidates, but high-resolution interferometric observations are necessary to distinguish between a non-magnetized and a strongly magnetized scenario. We computed dust emission maps from the RMHD models with RADMC-3D in bands 3, 4, 6, 7 and 9 of ALMA and built synthetic observations \citep{2012A&A...548A..39C} with the ALMA simulator \citep{2001sf2a.conf..569P} included in the GILDAS software package\footnote{\tt http://iram.fr/IRAMFR/GILDAS}. We considered four typical configurations of the full array, without ACA. The largest configurations allow to resolve the fragmentation scale of a few AU, but have a larger central hole in Fourier space, missing more of the large-scale flux than compact configurations. This flux loss increases at higher frequencies. Overall, observing FHSC candidates at 150 pc below 150 GHz with 500 m to 1 km baselines allows to resolve the fragmentation scale with a limited flux loss, thus suggesting an observing strategy for forthcoming ALMA proposals. ALMA's sensitivity allows for the observation of many such FHSC candidates in a single observing proposal. Details of these calculations and results can be found in \cite{2012A&A...545A..98C,2012A&A...548A..39C}.

\section{Perspectives}
Our work is currently limited to thermal dust continuum emission before the second collapse initiated by H$_2$ dissociation. An actively pursued extension is the inclusion of chemical processes in the collapsing core and the consequent search for line emission observational diagnostics. 

\bibliography{LevrierFrancois}

\end{document}